\documentclass[prb,twocolumn,superscriptaddress]{revtex4-2}

\usepackage{graphicx}
\usepackage{amsmath}
\usepackage{amssymb}
\usepackage{xcolor}
\usepackage{hyperref}
\usepackage{tikz}

\begin{document}

\title{Two-dimensional coherent spectrum of interacting spinons from matrix-product states}

\date{\today}

\author{Qi Gao}
\affiliation{Institute of Physics, Chinese Academy of Sciences, Beijing 100190, China}
\affiliation{University of Chinese Academy of Sciences, Beijing 100049, China}
\author{Yang Liu}
\affiliation{Institute of Physics, Chinese Academy of Sciences, Beijing 100190, China}
\affiliation{University of Chinese Academy of Sciences, Beijing 100049, China}
\author{Haijun Liao}
\email{navyphysics@iphy.ac.cn}
\affiliation{Institute of Physics, Chinese Academy of Sciences, Beijing 100190, China}
\affiliation{University of Chinese Academy of Sciences, Beijing 100049, China}
\affiliation{Songshan Lake Materials Laboratory, Dongguan, Guangdong 523808, China}
\author{Yuan Wan}
\email{yuan.wan@iphy.ac.cn}
\affiliation{Institute of Physics, Chinese Academy of Sciences, Beijing 100190, China}
\affiliation{University of Chinese Academy of Sciences, Beijing 100049, China}
\affiliation{Songshan Lake Materials Laboratory, Dongguan, Guangdong 523808, China}

\begin{abstract}
We compute numerically the second and third order nonlinear magnetic susceptibilities of an Ising ladder model in the context of two-dimensional coherent spectroscopy by using the infinite time-evolving block decimation method. The Ising ladder model couples a quantum Ising chain to a bath of polarized spins, thereby effecting the decay of spinon excitations. We show that its third order susceptibility contains a robust spinon echo signal in the weak coupling regime, which appears in the two-dimensional coherent spectrum as a highly anisotropic peak in the frequency plane. The spinon echo peak reveals the dynamical properties of the spinons. In particular, the spectral peak corresponding to the high energy spinons, which couple to the bath, is suppressed with increasing coupling, whereas those do not show no significant changes.
\end{abstract}

\maketitle

\section{Introduction}

The quest for quantum spin liquid, a highly entangled quantum phase of matter believed to exist in magnets, is a main endeavor of the modern condensed matter physics~\cite{Moessner2006,Balents2010,Savary2016,Zhou2017}. Among its many fascinating traits, the ability of hosting fractional excitations such as \emph{spinons} makes the quantum spin liquid a potential platform for the future quantum information technology. These excitations are made by breaking a magnon in half, and can carry quantum statistics that are neither boson or fermion.

Despite decades of experimental search, spinons so far remain elusive in two and three dimensional systems. A major obstacle toward its detection is that they have no sharp features in conventional spectroscopy. A magnon manifests itself in neutron or optical spectroscopy as a sharp resonance peak, whose center position and width are respectively tied to the energy and life time of the magnon. By contrast, the spinons, being fractional excitations, are always created in pairs in spectroscopy. As a result, the energy/momentum transfer from the probe to the sample may be distributed arbitrarily among the two spinons. Instead of a resonance peak, this process gives rise to a spectral continuum, which disguises the intrinsic properties of the spinons such as their lifetime.

Therefore, resolving the spectral continuum is an important first step toward the experimental identification and characterization of spinons. Recently, it has been suggested theoretically that \emph{two-dimensional coherent spectroscopy} may be used to tackle the problem~\cite{Wan2019}. The two-dimensional coherent spectroscopy probes the nonlinear optical susceptibilities of a sample using two phase coherent, ultrafast optical pulses~\cite{Mukamel1999,Hamm2011,Cundiff2013,Woerner2013,Lu2017,Parameswaran2020,Choi2020,Mahmood2021,Nandkishore2021,Li2021,Phuc2021,Fava2021,Gerken2022,Luo2022,Fava2022,Hart2022,Sim2022} The nonlinear responses are then plotted as a function of two frequency variables, resulting in a two-dimensional spectrum. Analytical calculations show that the $\chi^{(3)}$ response of the spinons contains an echo signal~\cite{Hahn1950,Kurnit1964} (dubbed ``spinon echo")~\cite{Wan2019}, which appears in the two-dimensional spectrum as a highly anisotropic peak. The width of the peak along the diagonal direction of the two-dimensional frequency plane matches the energy range of the spinon pair excitations akin to the continuum. Crucially, the width of the peak along the anti-diagonal is inversely proportional to the coherence time of spinon pairs, thus exposing the intrinsic property of the spinons and resolving the continuum.

The analysis of Ref.~\onlinecite{Wan2019} is carried out on the quantum Ising chain. In the ferromagnetic phase, this system hosts domain walls as the elementary excitations, which may be viewed as the one-dimensional analog of the spinons. Being exactly solvable, the spinons in this system are non-interacting, and their various nonlinear responses can be calculated straightforwardly. With interactions, the spinon may decay, thereby acquiring a finite lifetime. Ref.~\onlinecite{Wan2019} capture these effects phenomenologically by drawing analogy with the two-level systems and invoking the concept of coherent time ($T_2$ time) and the population time ($T_1$ time). While intuitive, this phenomenological treatment requires assessment from a more microscopic perspective. In particular, one may ask if the spinon echo is robust against spinon interactions and to what extent the phenomenology theory is reliable. 

In this work, we explore the effect of interactions on the spinon echo signal in \emph{Ising ladder}, a model tailored to exhibit spinon decay~\cite{Verresen2019}. In this model, the quantum Ising chain is coupled to another chain of polarized spins. The latter chain acts as a bath for the former so that a high energy spinons can dissipate its energy as it propagates. We calculate numerically the two-dimensional coherent spectrum by using the infinite time-evolving block decimation (iTEBD) method~\cite{Vidal2003,Vidal2004,Vidal2007}. Our scheme is inspired by the experimental protocol: We trigger the unitary evolution of the system by perturbing it with spatially uniform magnetic field pulses, and then monitor the ensuing magnetic response in real time, from which we extract the second order and third order nonlinear magnetic susceptibilities. This approach is straightforward to implement, numerically exact, and versatile. Yet, common to many numerical methods based on matrix product states, the accessible simulation time is limited by the growth of the entanglement entropy. Therefore, it complements nicely with analytical treatments~\cite{Hart2022}.

Our calculation shows that the two-dimensional spectrum exhibits robust spinon echo signal when the spinons are weakly coupled to the bath. Furthermore, the spinon echo signal decreases in magnitude as the pulse delay time increases, which is a direct manifestation of the decay of the spinons. We corroborate this interpretation by calculating the spinon spectral function and comparing it with the two-dimensional spectrum. Our work thus suggests the utility of the spinon echo beyond exactly solvable models.

The rest of this work is organized as follows. We describe the models and the numerical methods in Sec.~\ref{sec:model_method}. In Sec.~\ref{sec:results}, we present the main results. Finally, in Sec.~\ref{sec:discussion}, we discuss the limitations of the method used in this work and provide an outlook toward potential improvements and interesting open questions.

\section{Model and method \label{sec:model_method}}

\begin{figure}
\includegraphics[width=\columnwidth]{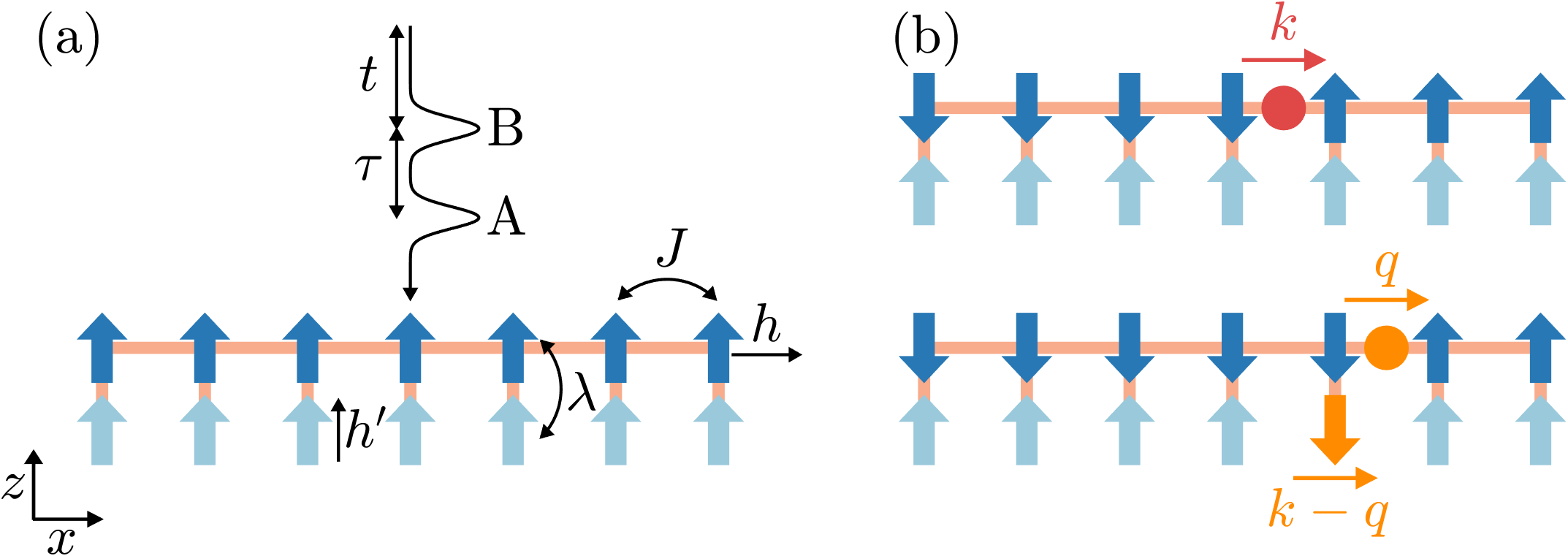}
\caption{(a) The Ising ladder consists of a quantum Ising chain (dark blue arrows) and a chain of polarized spins (light blue arrows), coupled by rung-wise interactions. The former chain is the system, whereas the latter acts as the bath. The two-dimensional coherent spectroscopy probes the nonlinear magnetic response of this system triggered by two successive, linearly polarized magnetic field pulses. (b) As a spinon propagates in the quantum Ising chain, it may excite a magnon in the bath chain and thereby dissipates its energy.}
\label{fig:model}
\end{figure}

The Hamiltonian of the Ising ladder model consists of three pieces (Fig.~\ref{fig:model}a):
\begin{subequations}
\begin{align}
H = H_S+H_B+V,
\end{align}

$H_S$ describes a quantum Ising chain, which is viewed as our system:
\begin{align} 
H_S =  -J\sum_{n}\sigma_{n}^z \sigma_{n+1}^z-h\sum_{n}\sigma_{n}^x,
\end{align}
where $n$ labels the lattice sites. $\sigma^{x,y,z}_n$ are the Pauli operators. $J>0$ is the exchange constant. $h>0$ is the transverse field. For simplicity, we rescale the energy such that $J+h \to 1$. When $J>h$, the quantum Ising chain is in the ferromagnetic phase and supports domain walls, or spinons, as fractional excitations.  

$H_B$ describes a chain of independent spins, which acts as a bath:
\begin{align}
H_B = - h' \sum_{n}\tau_{n}^z,
\end{align}
where $\tau^{x,y,z}_n$ are Pauli operators as well. $h'>0$ is an external field that polarizes the spins.

Finally, $V$ couples the bath to the system:
\begin{align}
V = -\lambda \sum_{n} \sigma_{n}^x\tau_{n}^x.
\end{align}
\end{subequations}
We focus on the perturbative regime $\lambda \ll 1$. In the absence of coupling, the bath spins are polarized, $\tau^z_n = 1$. As soon as $\lambda\neq 0$, a spinon propagates in the system chain may flip the bath spin and thereby transfers its energy to the bath (Fig.~\ref{fig:model}b). This mechanism gives rise to the desired spinon decay effect.

In the two-dimensional coherent spectroscopy experiment, two magnetic field pulses arrives at the sample, initially in equilibrium, at time $0$ and $\tau$ to induce magnetic responses, which are recorded at time $t$ after the arrival of the second pulse (Fig.~\ref{fig:model}a)~\cite{Mukamel1999,Hamm2011,Cundiff2013,Woerner2013,Lu2017}. In direct analogy of this procedure, we first put $H$ in its ground state $G$, and then apply two successive Dirac-$\delta$ pulses to the system. Here we set the pulses polarization to $x$ in order to excite pairs of spinon excitations. Mathematically, this is amount to applying a unitary rotation:
\begin{align}
|\psi'\rangle = e^{i\theta M^x} = e^{i\theta\sum_n \sigma^x_n} |\psi\rangle,
\end{align}
where $|\psi\rangle$ and $|\psi'\rangle$ are respectively the state before and after the pulse. $\theta$ is the total area under the pulse. Note we assume only the system spin ($\sigma$ spins) couple to the pulse. The magnetization is thus given by:
\begin{align}
m^x (t,\tau,\theta_a,\theta_b) = \langle \psi | \sigma^x_n |\psi \rangle,
\end{align}
where
\begin{align}
|\psi \rangle = e^{-itH}e^{i\theta_b M^x}e^{-i\tau H} e^{i\theta_a M^x} |G\rangle.
\end{align}
Here, $\theta_{a,b}$ are respectively the strength of the first and second pulses. The magnetization is independent of the site $n$ we choose to measure thanks to the translation invariance.

$m^x(t,\tau,\theta_a,\theta_b)$ comprises of contributions from both linear and various order nonlinear susceptibilities. To extract the nonlinear susceptibilities, which are our focus here, we take its derivative with respect to $\theta_{a,b}$:
\begin{subequations}
\begin{align}
 \chi_{xxx}^{(2)}(t,t+\tau) &= \left. \frac{\partial^2 m^x (t,\tau,\theta_a,\theta_b)}{\partial\theta_a \partial \theta_b} \right|_{\theta_{a,b}=0};
 \\
 \chi_{xxxx}^{(3)}(t,t,t+\tau) &= \left. \frac{\partial^3 m^x (t,\tau,\theta_a,\theta_b)}{\partial\theta_a \partial \theta_b^2} \right|_{\theta_{a,b}=0};
 \\
 \chi_{xxxx}^{(3)}(t,t+\tau,t+\tau) &=\left. \frac{\partial^3 m^x (t,\tau,\theta_a,\theta_b)}{\partial\theta_a^2 \partial \theta_b} \right|_{\theta_{a,b}=0}.
\end{align}
\end{subequations}
In particular, $\chi^{(3)}_{xxxx}(t,t,t+\tau)$ corresponds to the third order process where the first pulse couples linearly to the system whereas the second pulse couples quadratically. For $\chi^{(3)}_{xxxx}(t,t+\tau,t+\tau)$, the role of the first and second pulses are switched. Performing a Fourier transform over the time variables $t$ and $\tau$ yields the two-dimensional coherent spectrum from each of these nonlinear susceptibilities.

In practice, we approximate the derivative by a finite difference. We find the \emph{central} difference schemes, which are symmetric with respect to $\theta_{a}$ and $\theta_b$, to be optimal. The difference scheme for $\chi_{xxx}^{(2)}$ is given by:
\begin{align}
\chi_{xxx}^{(2)} =& \frac{1}{4 \theta^2}[m^x (\theta,\theta)-m^x (\theta,-\theta)
\nonumber\\
&-m^x (-\theta,\theta)+m^x (-\theta,-\theta)],
\end{align}
where we have suppressed the time variables for brevity. Therefore, we need to calculate $m^x$ at four different pulse strengths. We find choosing $\theta = 0.01$ is sufficient accurate for our purpose. The difference schemes for the other susceptibilities may be obtained in the same vein.

The remaining task is thus to find the ground state $|G\rangle$ and evolve it in real time. To this end, we use the imaginary time iTEBD algorithm \cite{Vidal2003,Vidal2004,Vidal2007} to find the ground state, and then the real time version to carry out the time evolution. The maximal bound dimension is $D = 1200$, resulting in a typical truncation error on the order of $10^{-5} \sim 10^{-8}$ upon terminating the simulation. For both procedures, we use the second order Trotter-Suzuki decomposition. We find a moderate Trotter time $\Delta t = 0.25$ is sufficiently accurate for our purpose while significantly reducing the computational time.

\begin{figure}
\includegraphics[width=\columnwidth]{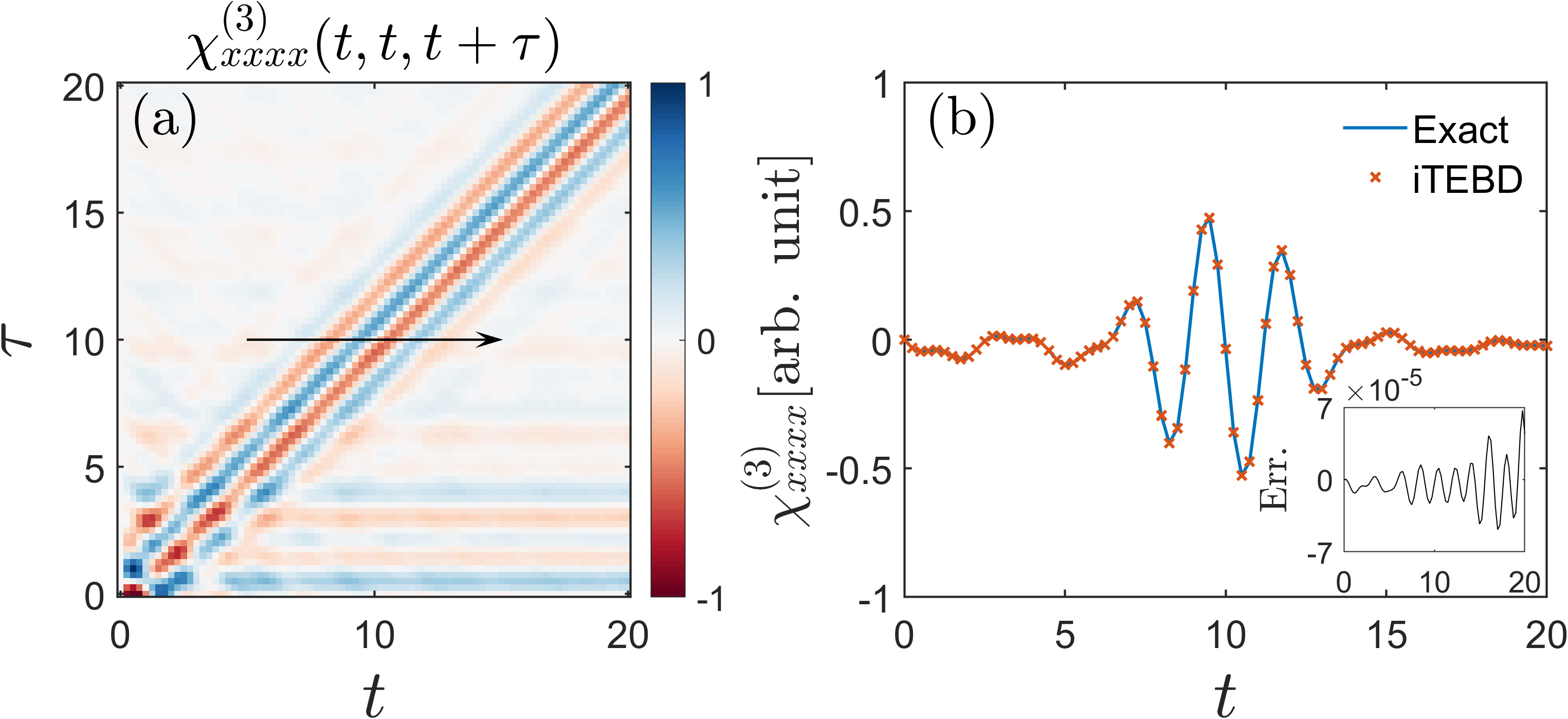}
\caption{(a) The nonlinear susceptibilities $\chi_{xxxx}^{(3)}(t,t,t+\tau)$ of the quantum Ising chain as a function of $t$ and $\tau$. $J=0.7$ and $h=0.3$.  (b) iTEBD and exact result for $\chi_{xxxx}^{(3)}(t,t,t+\tau)$ as a function of $t$ at fixed $\tau = 10$ (marked by the arrow in (a)). The inset shows the error of the iTEBD result.}
\label{fig:benchmark}
\end{figure}

We benchmark our method with the quantum Ising chain ($\lambda=0$), whose nonlinear responses can be found analytically. Fig.~\ref{fig:benchmark}a shows the third order nonlinear magnetic susceptibility $\chi^{(3)}_{xxxx}(t,t,t+\tau)$ as a function of $t$ and $\tau$ for the model parameters $J=0.7$ and $h=0.3$. The data are scaled such that the maximum value is 1. Our result agrees well with the exact result, where we perform the Trotterized time evolution analytically. To facilitate a quantitative comparison, we plot in Fig.~\ref{fig:benchmark}b the data along a $t$-scan (marked as black arrow in the panel a) as well as the analytical result. Note the data from iTEBD and from the analytical solution are scaled by the \emph{same} factor. We find the error is on the order of $10^{-5}$. This excellent agreement demonstrates the reliability of our method.

\begin{figure}
\includegraphics[width=1\linewidth]{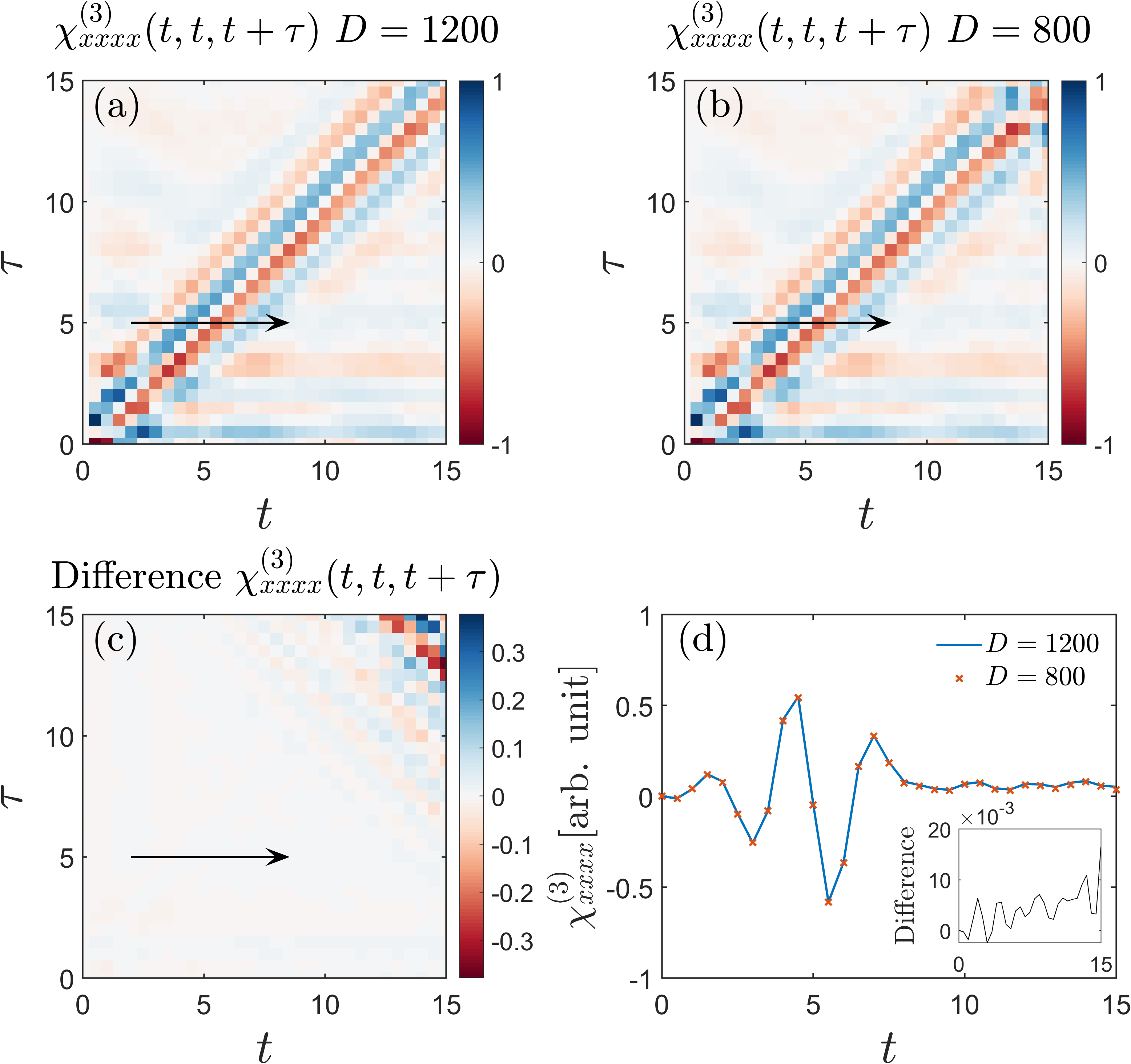}
\caption{(a) The nonlinear susceptibilities $\chi_{xxxx}^{(3)}(t,t,t+\tau)$ as a function of $t$ and $\tau$, obtained with the bond dimension $D = 1200$ and the Trotter time $\Delta t = 0.25$. The model parameters $J=0.65$, $h=0.35$ ,$h'=0.4$, and $\lambda=0.1$. (b) The same as (a) but with the bond dimension $D=800$ and the Trotter time $\Delta t = 0.05$. Note both data are scaled by the same factor. (c) Difference between the data shown in (a) and (b). (d) Comparison of the data in (a) and (b) as a function of $t$ and fixed $\tau = 5$ (marked by the arrow in (a)). The inset shows the difference.}
\label{fig:convergence}
\end{figure}

As the Ising ladder model is no longer solvable when $\lambda\neq0$,  we test the convergence of the iTEBD algorithm by varying the bond dimension $D$ and the trotter time $\Delta t$. Fig.~\ref{fig:convergence} compares the nonlinear susceptibility $\chi^{(3)}_{xxxx}(t,t,t+\tau)$ for a representative set of model parameters, obtained with $D = 1200$ and $\Delta t = 0.25$ (a) and $D = 800$, $\Delta t = 0.05$ (b). Note the data in these two plots are scaled by the \emph{same} factor. We find the error grows as $t$ or $\tau$ increases  (Fig.~\ref{fig:convergence}c), reflecting the growth of the entanglement entropy. Nevertheless, the data clearly have converged for $t,\tau$ up to 10 at bond dimension $D=800$. Indeed, the error along the cut marked as the black arrow is on the order of $10^{-3}$ (Fig.~\ref{fig:convergence}d).

Finally, we characterize the spinon dynamics directly by using spinon spectral function. To this end, we define an operator:
\begin{align}
d^\dagger_{k} &\equiv \frac{1}{\sqrt{L}} \sum^L_{n=1} e^{-ik(n-1)} (c_n+c^\dagger_n)
\nonumber\\
& = \frac{1}{\sqrt{L}} \sum^L_{n=1} e^{-ik(n-1)} \prod^{n-1}_{m=1} (-\sigma_m^x) \cdot \sigma^z_n.
\end{align}
Here, $c_n$ and $c^\dagger_n$ are Jordan-Wigner fermion operators. In the second line, we have inserted their explicit expression in terms of Pauli operators. $L$ is the length of the chain. It's straightforward to verify that, in the thermodynamic limit, applying this operator to the ground state of the quantum Ising chain($\lambda=0$) yields a spinon excitation with momentum $k$ up to a phase,
\begin{align}
d^\dagger_{k} |G\rangle = |k\rangle,\quad (\lambda=0).
\end{align}
This observation motivates us to define the Green's function as follows:
\begin{align}
iG(k,t) = \theta(t)\langle G| [d^{\phantom{\dagger}}_k(t),d^\dagger_k(0)]_+ |G\rangle,
\end{align}
where $[\cdot,\cdot]_+$ represents the anti-commutator. The spinon spectral function is then obtained by its Fourier transform:
\begin{align}
A(k,\omega) = -\frac{1}{\pi}\mathrm{Im}G(k,\omega).
\end{align} 

We compute $G(k,t)$ by using the TEBD algorithm \cite{Vidal2003,Vidal2004} on a finite size chain with open boundary conditions. This is because the Jordan-Wigner transform is most conveniently expressed with open boundary condition. Specifically, we recast $d^\dagger_k$ as a $D=2$ matrix-product operator (MPO)~\cite{Wu2020}:
\begin{subequations}
\begin{equation}
    d^\dagger_k =
        \begin{tikzpicture}[baseline = (X.base),every node/.style={scale=0.75},scale=.55]
            \draw (2,1.5) circle (0.5);
            \draw (2,1.5) node (X) {$\nu_L$};
            \draw (2.5,1.5) -- (3,1.5); 
            \draw[rounded corners] (3,2) rectangle (4,1);
            \draw (3.5,1.5) node {$M$};
            \draw (4,1.5) -- (5,1.5);
            \draw[rounded corners] (5,2) rectangle (6,1);
            \draw (5.5,1.5) node {$M$};
            \draw[dotted] (6,1.5) -- (7,1.5); 
            \draw[ rounded corners] (7,2) rectangle (8,1);
            \draw (7.5,1.5) node {$M$};
            \draw (8,1.5) -- (9,1.5); 
            \draw[rounded corners] (9,2) rectangle (10,1);
            \draw (9.5,1.5) node {$M$};
            \draw (10,1.5) -- (10.5,1.5);
            \draw (11,1.5) circle (0.5);
            \draw (11,1.5) node {$\nu_R$};
            \draw (3.5,1) -- (3.5,.5); \draw (5.5,1) -- (5.5,.5);
            \draw (7.5,1) -- (7.5,.5); \draw (9.5,1) -- (9.5,.5);
            \draw (3.5,2) -- (3.5,2.5); \draw (5.5,2) -- (5.5,2.5);
            \draw (7.5,2) -- (7.5,2.5); \draw (9.5,2) -- (9.5,2.5);
        \end{tikzpicture},
\end{equation}
where
\begin{align}
M = \begin{bmatrix}
\mathbb{I} & 0\\
\sigma^z & -e^{-ik}\sigma_x \\
\end{bmatrix};
\,
\nu_L = \begin{bmatrix}
0 & \mathbb{I}
\end{bmatrix};
\,
\nu_R = \begin{bmatrix}
\mathbb{I} \\
0
\end{bmatrix}.
\end{align}
\end{subequations}
This MPO representation fits nicely with the TEBD algorithm. For the real time evolution, we use the maximal bound dimension $D=100$ and Trotter time $\Delta t = 0.02$.

As the spinon spectral function is calculated for a finite size chain, we use the numerical link-cluster expansion \cite{Tang2013} to reduce the finite size effect. We find that using two chains with size $L=60$ and $L=62$ are sufficient. For the Fourier transform, we extrapolate the time domain data $G(k,t)$ to larger $t$ by using the linear prediction technique \cite{White2008,Barthel2009}, which yields better frequency resolution for the spectral function.

\section{Results \label{sec:results}}

In this section, we present the results on the two-dimensional coherent spectrum of the Ising ladder. We first demonstrate that the spinon does decay by analyzing its spinon spectral function. We then proceed to present the nonlinear magnetic responses in both the time and the frequency domain, and relate their features to the dynamical properties of spinons.

\begin{figure}
\includegraphics[width = \columnwidth]{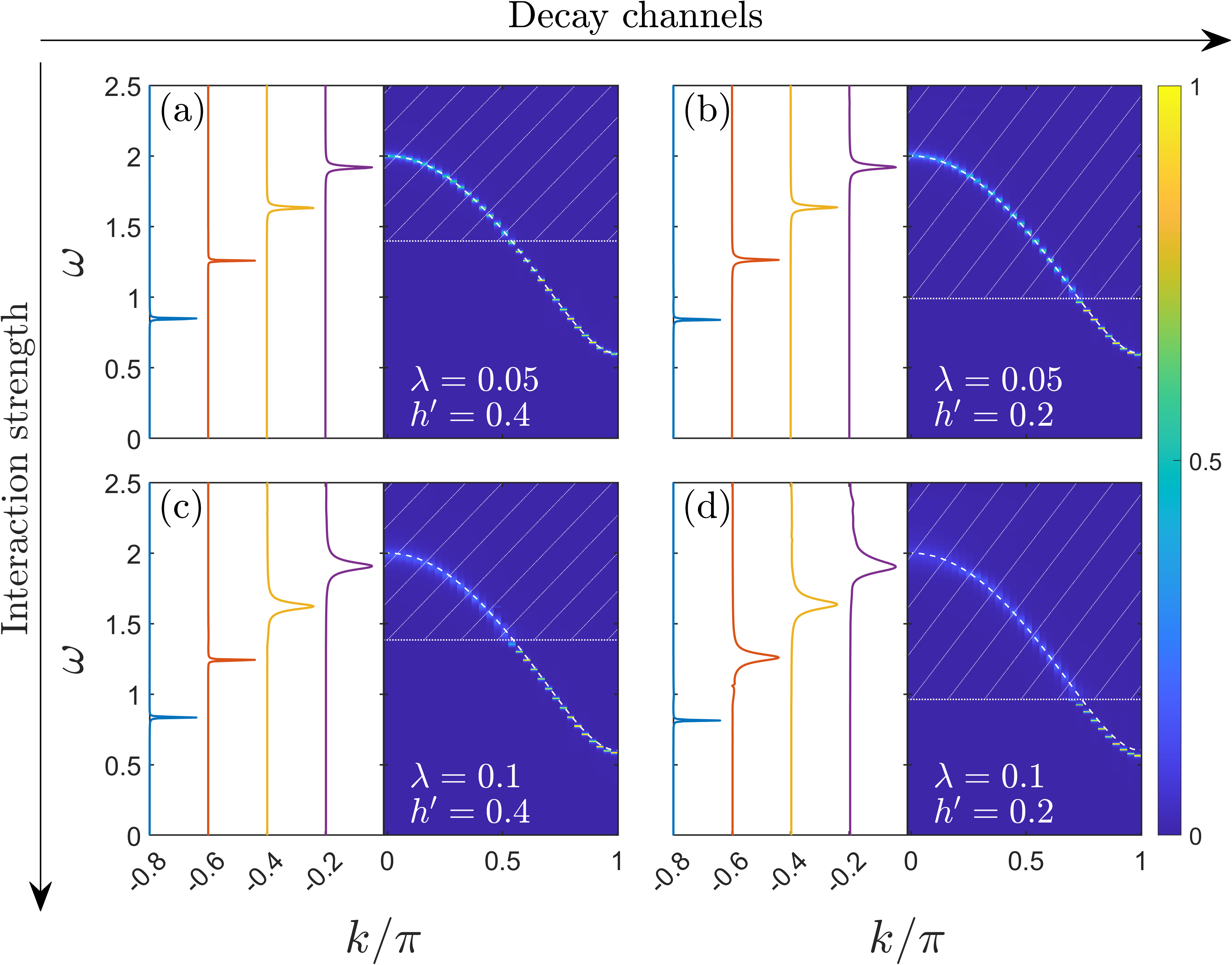}
\caption{(a) Spinon spectral function of the Ising ladder. The model parameters $J=0.65$, $h=0.35$, $h'=0.4$, and $\lambda = 0.05$. The right half of the panel shows the spectrum $A(k,\omega)$ as a function of  spinon momentum $k$ and the spinon energy $\omega$. Only the $\omega>0$ part is shown. The white dashed line marks the dispersion relation of the free spinon. The hatched area demarcates the region where the spinon decay is kinematically allowed. The left half of the panel shows the line shape of the spectral function at four representative momentum points $k/\pi = -0.2,-0.4,-0.6,-0.8$. Note $A(k,\omega)$ is an even function of $k$. (b) Same as (a) but with $h' = 0.2$ and $\lambda = 0.05$. (c)  $h' = 0.4$ and $\lambda = 0.1$. (d)  $h' = 0.2$ and $\lambda = 0.1$. }
\label{fig:dispersion}
\end{figure}

To set the stage, we analyze the dynamics of a single spinon in the Ising ladder. In this model, a spinon with momentum $k$ may decay into a state with lower energy at momentum $q$ and an additional magnon with momentum $k-q$ in the bath (Fig.~\ref{fig:model}b). The conservation of energy requires that:
\begin{align}
\epsilon_k = \epsilon_q + \omega_{k-q} = \epsilon_q + 2h',
\end{align}
where $\epsilon_k$ and $\omega_{k-q}$ are respectively the dispersion relation of the spinon and magnon. In the second equality, we have used the fact that the magnon in the bath has a constant energy $2h_b$, independent of the momentum. The above relation immediately implies the following kinematic constraint on the spinon decay:
\begin{align}
\epsilon_k > \Delta+2 h' \equiv \epsilon_{th},
\end{align}
where we have defined the decay threshold energy $\epsilon_{th}$. $\Delta = 2|J-h|$ is the energy gap for the spinon excitation. In the ensuing discussion, we set $J=0.65$ and $h=0.35$, which yields $\Delta = 0.6$. Spinons with energy above the threshold may decay and, as a result, have finite lifetime, whereas those below the threshold remain essentially free particles.

Fig.~\ref{fig:dispersion}a presents the spinon spectral function for $h' = 0.4$, and $\lambda = 0.05$. With such a choice of parameters, the decay threshold $\epsilon_{th} = 1.4$. As the momentum $k$ goes through the Brillouin zone, the position of the center of the spectral peak essentially follows the dispersion relation of the free spinon. However, the spectral peak broadens once it is above the decay threshold $\epsilon_{th}$. Specifically, the line width of the peak is resolution limited below $\epsilon_{th}$, and finite above it. These results are in excellent agreement with the preceding analysis. 

We further explore the physics of spinon decay by varying the parameters $h'$ and $\lambda$. On one hand, decreasing $h'$ from 0.4 to 0.2 reduces the threshold, and, accordingly, we find the spinons a larger portion of the Brillouin zone can decay (Fig.~\ref{fig:dispersion}b). On the other hand, increasing the value of $\lambda$ does not reduce the decay threshold but enhances the decay rate. As a result, dialing up $\lambda$ from 0.05 to 0.1, we find the decay threshold is identical but the line width is now significantly larger above the threshold (Fig.~\ref{fig:dispersion}c). Finally, simultaneously reducing $h'$ and increasing $\lambda$ lowers the decay threshold \emph{and} increases the decay rate (Fig.~\ref{fig:dispersion}d).

\begin{figure}
\includegraphics[width = \columnwidth]{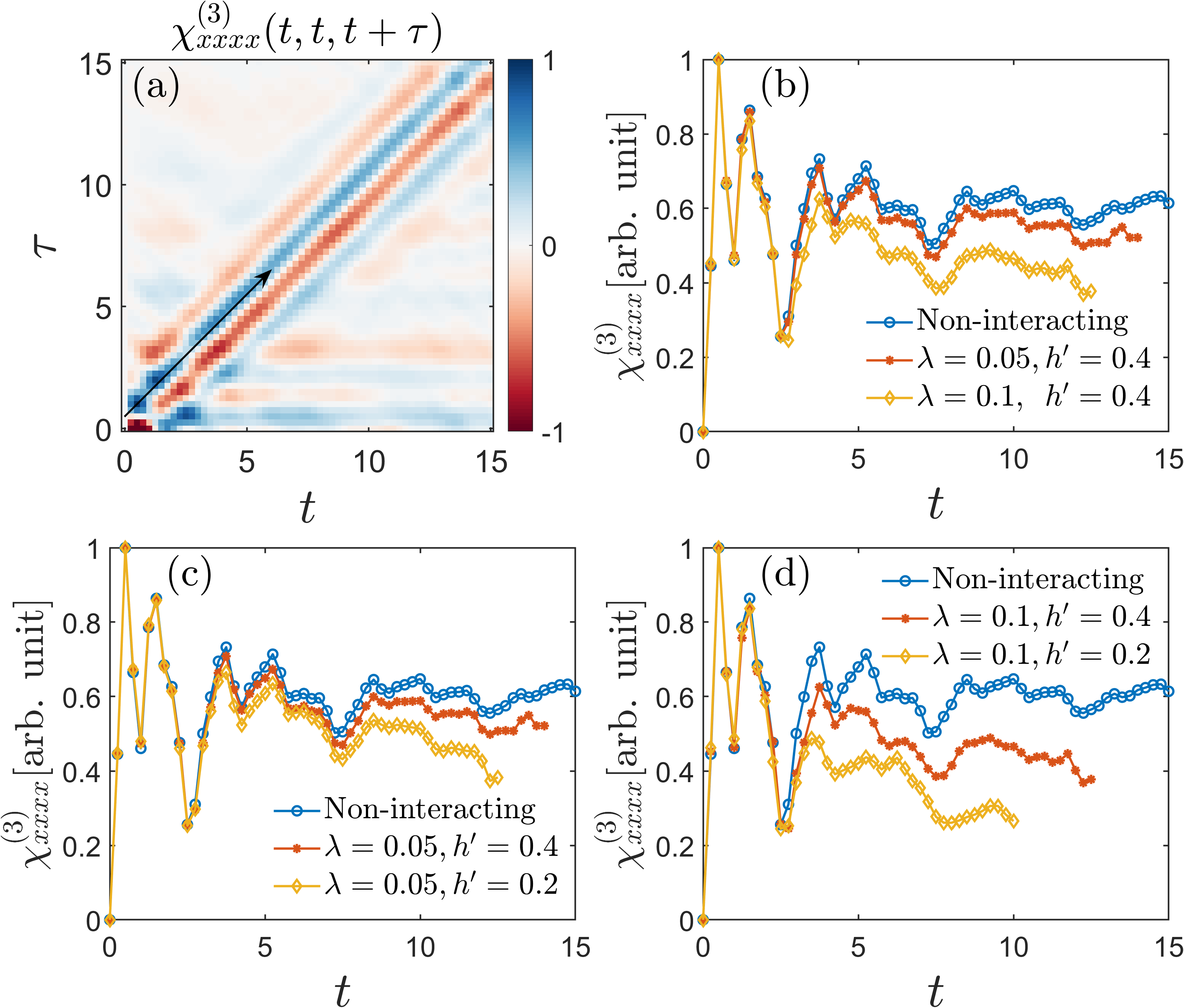}
\caption{(a) Nonlinear magnetic susceptibility $\chi^{(3)}_{xxxx}(t,t,t+\tau)$ of the Ising ladder as a function of $t$ and $\tau$. The model parameters $J = 0.65$, $h = 0.35$, $h' = 0.4$, $\lambda = 0.1$. The data are scaled such that the maximum value is 1. (b) The profile of the nonlinear susceptibility along the diagonal direction (shown as the black arrow in panel (a)) with increasing coupling $\lambda$ and fixed $h' = 0.4$. (c)(d) Similar to panel (b) but with decreasing $h'$ and fixed $\lambda = 0.05$ (c) and $\lambda = 0.1$ (d).}
\label{fig:bba_time}
\end{figure}

Having demonstrated spinon decay, we now investigate its signature in the two-dimensional coherent spectrum. In the phenomenological theory, the spinon pairs excited by the magnetic field pulses are mapped to two-level systems. The pairs with different momenta $\pm k$ are treated as dynamically decoupled. Moreover, the interaction effects are subsumed in the phenomenological phase coherent time ($T_2$ time) and the population time ($T_1$ time) of the two-level systems. The calculation then shows that the nonlinear magnetic susceptibility $\chi^{(3)}_{xxxx}(t,t,t+\tau)$ can be used to probe the phase coherence time (the $T_2$ time) of the spinon excitations. Specifically~\cite{Wan2019},
\begin{align}
\chi^{(3)}_{xxxx}(t,t,t+\tau) &= -\frac{32}{\pi}\int^\pi_0 dk\sin^4\theta_k \sin[2\epsilon_k (t-\tau)]
\nonumber\\
& \times e^{-(t+\tau)/T_{2,k}}+\cdots.
\end{align}
Here, the integration is over contributions from all spinon pairs with momenta $\pm k$. $2\epsilon_k$ is their energy, whereas $T_{2,k}$ is the phenomenological phase coherence time. $\sin\theta_k$ is an optical matrix element defined implicitly through $\tan\theta_k = J\sin k/(J\cos k + h)$. We have dropped the terms that quickly decay in late time. Setting $t-\tau = \delta$ which is a constant, we obtain:
\begin{align}
\chi^{(3)}_{xxxx} \overset{t-\tau=\delta}{\sim} \int^\pi_0 dk\sin^4\theta_k \sin(2\epsilon_k \delta ) e^{-2t/T_{2,k}}.
\end{align}
In other words, $\chi^{(3)}_{xxxx}(t,t,t+\tau)$ would persist along the line $t-\tau = \delta$,  which is the spinon echo signal. The slow decay of the spinon echo signal as a function of $t$ is a direct manifestation of the phase coherence time of the spinon pairs ($T_{2,k}$).

Fig.~\ref{fig:bba_time}a presents the nonlinear magnetic susceptibility $\chi^{(3)}_{xxxx}(t,t,t+\tau)$ as a function of the time variables $t$ and $\tau$ for $h’=0.4$, and $\lambda=0.1$. The spinon echo signal appears as the feature that persists along the diagonal direction of the plot as predicted by the phenomenological theory. The pronounced echo signal reveals the highly coherent spinon excitations in this system.

Fig.~\ref{fig:bba_time}b shows the profile of the nonlinear magnetic susceptibility $\chi^{(3)}_{xxxx}(t,t,t+\tau)$ along the diagonal direction with increasing $\lambda$ and fixed $h' = 0.4$. For the free Ising chain ($\lambda=0$), the echo approaches a constant as $t$ increases. Dialing in the coupling $\lambda$ leads to the decay of the spinon excitations, and, as a result, suppresses the echo signal. Meanwhile, reducing the value of $h'$ also suppresses the echo signal (Fig.~\ref{fig:bba_time}c\&d), which is naturally understood as the result of the lower decay threshold $\epsilon_{th}$, and, in turn, more decaying spinon modes. 

While the magnitude of the echo signal decreases in time, we stress that it is \emph{not} expected to disappear completely as $t\to \infty$ in our model --- There remain free spinon modes below the decay threshold (Fig.~\ref{fig:dispersion}), which would give rise to echo signal that persists in time. We shall come back to this point momentarily.

\begin{figure}
\includegraphics[width = \columnwidth]{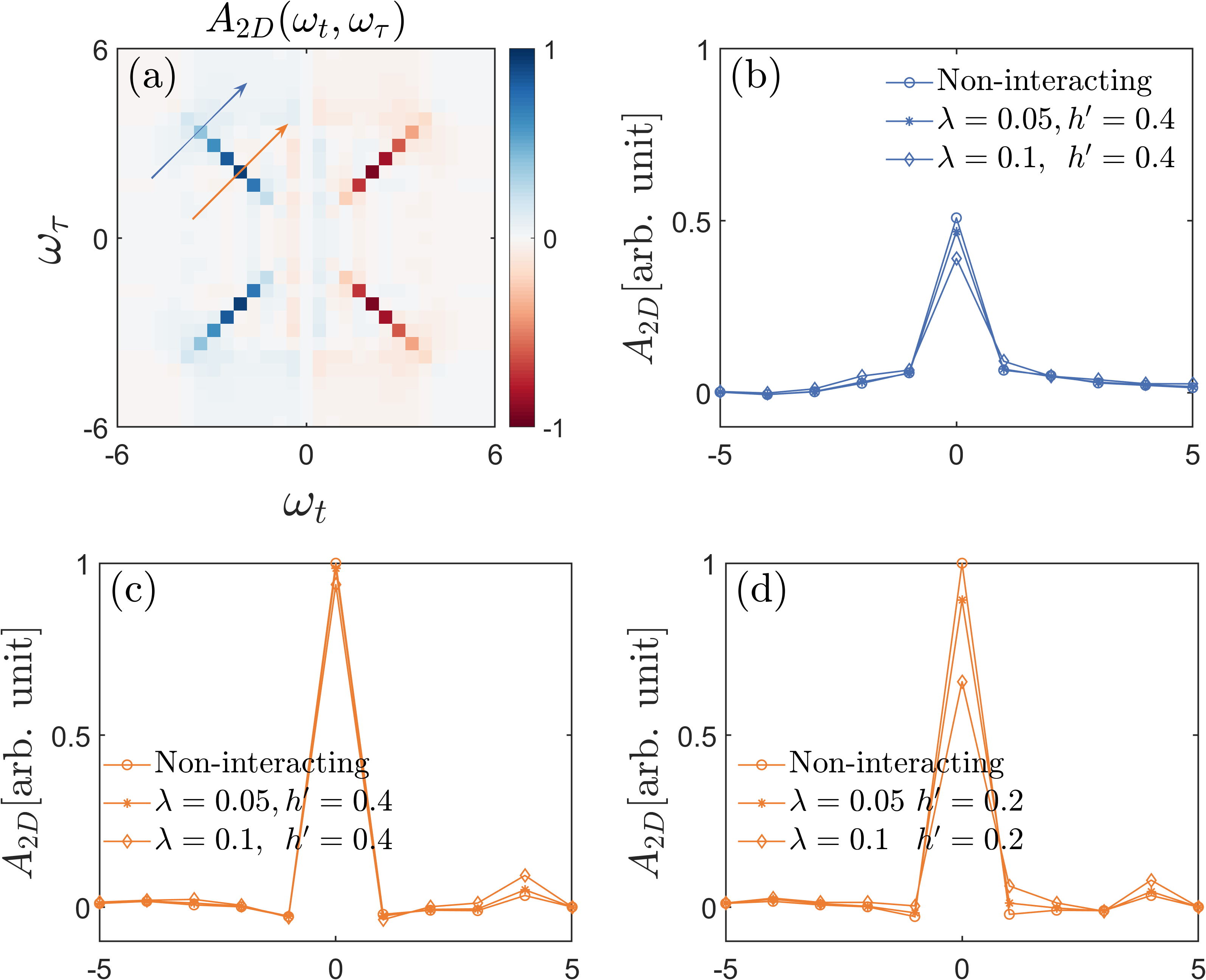}
\caption{(a) Phase-untwisted two-dimensional coherent spectrum from the nonlinear magnetic response $\chi^{(3)}_{xxxx}(t,t,t+\tau)$. The model parameters $J=0.65$, $h=0.35$, $h' = 0.4$, and $\lambda = 0.1$. (b) The profile of the spectrum along the cut marked as the blue arrow in panel (a) for increasing $\lambda$ and fixed $h' = 0.4$. (c) The profile of the spectrum along the cut marked as the orange arrow in panel (a) for increasing $\lambda$ and fixed $h'=0.4$. (d) Similar to (c) but for $h' = 0.2$.}
\label{fig:bba_fft}
\end{figure}

Fourier-transforming the time domain data in Fig.~\ref{fig:bba_time}a produces the two-dimensional coherent spectrum, which contains more information (Fig.~\ref{fig:bba_fft}a). It is convenient to use the phase untwisting trick, namely symmetrizing the imaginary part of the Fourier-transform with respect to $\omega_\tau$~\cite{Hamm2011,Khalil2003}. The phenomenological theory shows the spectrum is given by~\cite{Wan2019}:
\begin{align}
A_{2D}(\omega_t,\omega_\tau) &\equiv \mathrm{Im}\chi^{(3)}_{xxxx}(\omega_t,\omega_\tau)+\mathrm{Im}\chi^{(3)}_{xxxx}(\omega_t,-\omega_\tau) 
\nonumber\\
& = -32\pi \int^\pi_0 dk \sin^4\theta_k [f_k(\omega_\tau)+f_k(-\omega_\tau)] 
\nonumber\\
& \times [f_k(\omega_t)-f_k(-\omega_t)]+\cdots,
\end{align}
where $f_k(\omega)$ is the Lorentzian function:
\begin{align}
f_k(\omega) = \frac{1}{\pi}\frac{T_{2,k}}{1+[(\omega-2\epsilon_k)T_{2,k}]^2}.
\end{align}
The spinon echo peaks thus appear as the ``streaks" in the diagonal directions of the frequency plane. The broad width of the peak along the diagonal direction reflects the energy range of a pair of spinons with zero total momentum, i.e. twice of the spinon band width. More importantly, the line width of the spinon echo peak along the anti-diagonal direction is expected to reveal the coherence time of the spinon pairs --- its anti-diagonal width measured at frequency $\pm\omega_\tau =  \pm\omega_t = 2\epsilon_k$ is proportional to $1/T_{2,k}$. In what follows, we examine the profile of the two-dimensional coherent spectrum along various anti-diagonal cuts. 

Fig.~\ref{fig:bba_fft}b shows the spinon echo peak along the cut marked as the blue arrow in Fig.~\ref{fig:bba_fft}a, which corresponds to the spinon pair with energy $2\epsilon_k \approx 3.4$. Recall that the decay threshold for a single spinon is at $\epsilon_{th} = 1.4$. This pair thus lie above the threshold and are expected to have finite lifetime. Here, we are unable to resolve the line width of the peak due to limited simulation time. Nevertheless, we observe that the height of the peak does decrease monotonically as the coupling to the bath $\lambda$ increases, suggesting that the spinon pair at this energy do couple to the bath and thus may decay.

By contrast, the spinon echo peak cut along the orange arrow in Fig.~\ref{fig:bba_fft}c corresponds to the spinon pair with energy $2\epsilon_k = 2.1$, which lie below the decay threshold. As a result, the peak does not show significant change as the coupling $\lambda$ increases (Fig.~\ref{fig:bba_fft}c). If we set $h' = 0.2$ and thereby reduce the decay threshold to $\epsilon_{th} = 1$, the height of the peak now decreases with increasing $\lambda$ (Fig.~\ref{fig:bba_fft}d), suggesting that this pair of spinons now couple to the bath, and, therefore, may decay. 

\begin{figure}
\includegraphics[width = \columnwidth]{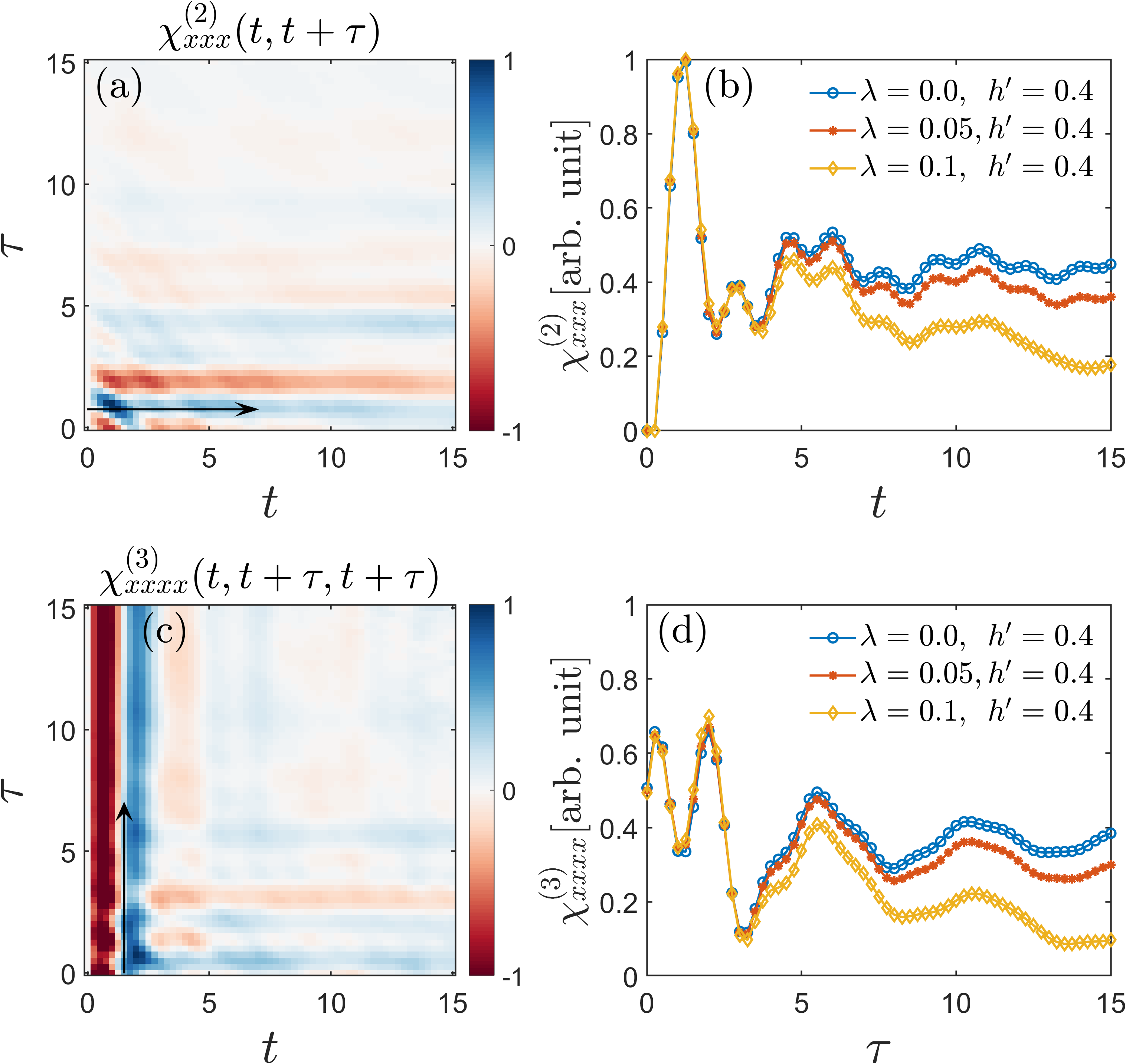}
\caption{(a) Nonlinear magnetic susceptibility $\chi^{(2)}_{xxx}(t,t+\tau)$ as a function of $t$ and $\tau$. The model parameters $J = 0.65$, $h = 0.35$, $h' = 0.4$, and $\lambda = 0.1$. The data are scaled such that the maximum value is 1. (b) The nonlinear susceptibility as a function of $t$ and constant $\tau = 0.75$ (black arrow in panel (a)) for various values of $\lambda$ and fixed $h' = 0.4$. (c)(d) Same as (a)(b) but for the nonlinear magnetic susceptibility $\chi^{(3)}_{xxxx}(t,t+\tau,t+\tau)$.}
\label{fig:t1_time}
\end{figure}

Finally, we discuss the other nonlinear magnetic susceptibilities that can be probed by the two-dimensional coherent spectroscopy, namely $\chi^{(2)}_{xxx}(t,t+\tau)$ and $\chi^{(3)}_{xxxx}(t,t+\tau,t+\tau)$. The phenomenological theory suggests that these two nonlinear responses can be used to detect the population time ($T_1$ time) of the spinon pairs. For the second order susceptibility, the phenomenological theory predicts that~\cite{Wan2019}:
\begin{align}
\chi^{(2)}_{xxx}(t,t+\tau) &= \frac{16}{\pi}\int^\pi_0 dk \sin^2\theta_k \cos\theta_k  \cos(2\epsilon_k\tau)  e^{-\tau/T_{2,k}}
\nonumber\\
& \times e^{-t/T_{1,k}} + \cdots,
\label{eq:ab_time}
\end{align}
where $T_{1,k}$ and $T_{2,k}$ are respectively the phenomenological population time and coherence time of the spinon pair with momenta $\pm k$. $\sin\theta_k$ and $\cos\theta_k$ are optical matrix elements defined before. We have omitted the terms that are washed out by dephasing at late times. Eq.~\ref{eq:ab_time} shows that $\chi^{(2)}_{xxx}$ would decrease in magnitude as a function of $t$ and fixed $\tau$, which is a manifestation of the population time. Likewise,
\begin{align}
\chi^{(3)}_{xxxx}(t,t+\tau,t+\tau) &= -\frac{64}{\pi} \int^\pi_0 dk \sin^4\theta_k \sin(2\epsilon_k t)
\nonumber\\
& \times e^{-t/T_{2,k}} e^{-\tau/T_{1,k}}+\cdots,
\end{align} 
which indicates that this nonlinear signal will decay with $\tau$ with fixed $t$.

We now compare the numerical results against the phenomenological theory. Fig.~\ref{fig:t1_time}a shows $\chi^{(2)}_{xxx}(t,t+\tau)$ as a function of $t$ and $\tau$. The model parameters $h' = 0.4$, and $\lambda = 0.1$.  The data decreases in magnitude as a function of $t$ and constant $\tau$ (Fig.~\ref{fig:t1_time}b). Furthermore, increasing $\lambda$ leads to larger suppression at late time. Fig.~\ref{fig:t1_time}c\&d show $\chi^{(3)}_{xxxx}(t,t+\tau,t+\tau)$ as a function of $t$ and $\tau$ with the same set of model parameters. We find its magnitude decreases with $\tau$ while holding $t$ fixed. All of these features are qualitatively consistent with the phenomenological theory.

\section{Discussion \label{sec:discussion}}

\begin{figure}
\includegraphics[width = \columnwidth]{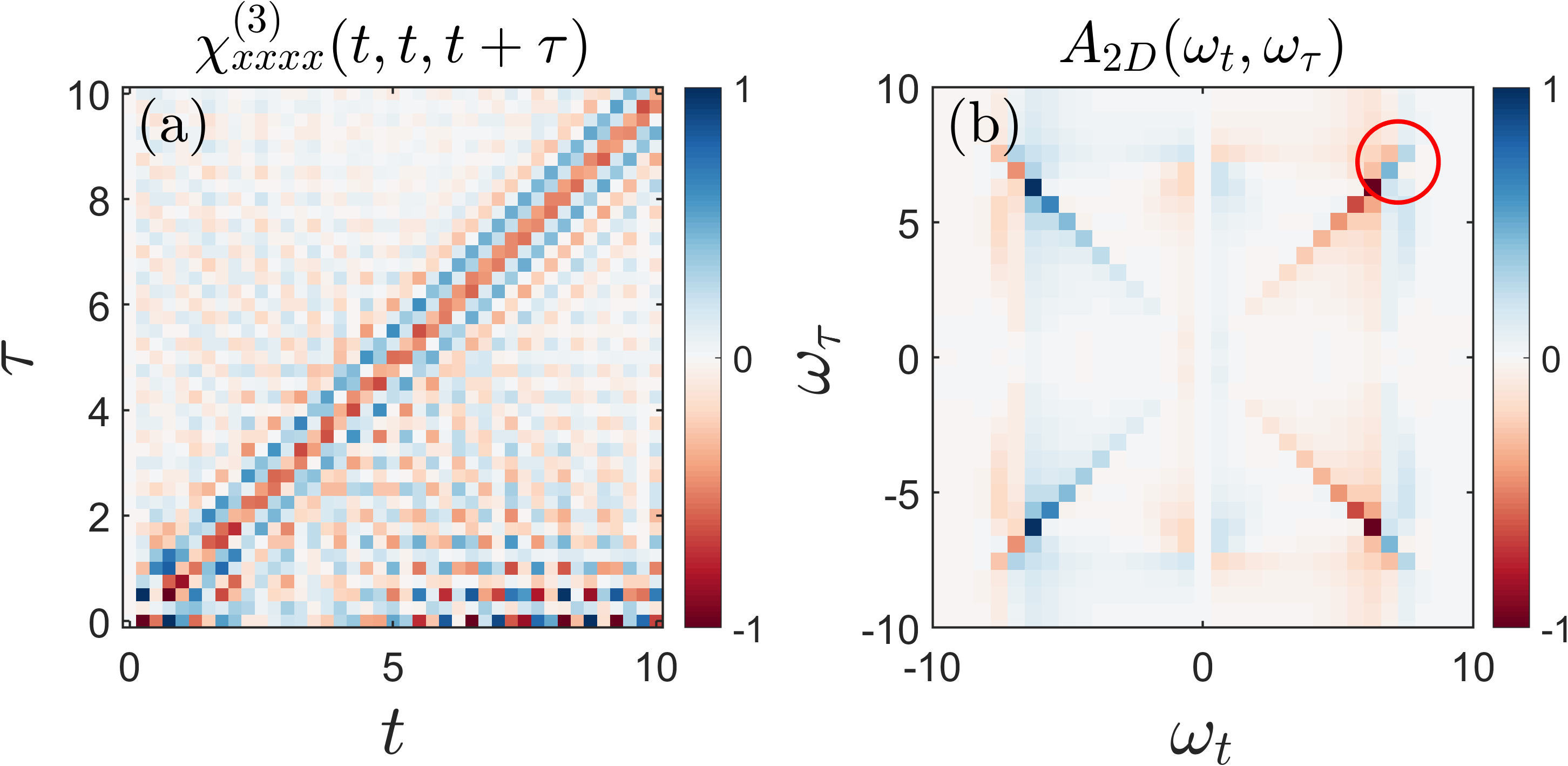}
\caption{(a) Nonlinear magnetic susceptibility $\chi^{(3)}_{xxxx}(t,t,t+\tau)$ as a function of $t$ and $\tau$ for a perturbed Ising mode. The model parameters $J=1$, $h=0.4$, $J_2 = -0.48$, and $\lambda = 0.2$. The data are scaled such that the maximum value is 1. (b) The phase-untwisted two-dimensional coherent spectrum. The red circle highlights the feature arises from the oscillatory signal in the time-domain data.}
\label{fig:perturbed_model}
\end{figure}

To summarize, we have calculated numerically the nonlinear magnetic susceptibilities of the Ising ladder model. This model is tailored to exhibit the decay of spinon excitations by coupling them to a bath. In the weak coupling regime, we find $\chi^{(3)}_{xxxx}(t,t,t+\tau)$ exhibits the spinon echo signal  that reveals the highly coherent spinon excitations. The magnitude of the spinon echo signal decreases as the pulse delay time increases, which reflects the decay of the spinon excitations. Furthermore, the two-dimensional coherent spectrum, obtained by Fourier-transforming $\chi^{(3)}_{xxxx}(t,t,t+\tau)$, suggests that the high energy spinon pairs can decay whereas the low energy pairs do not. In addition, the susceptibilities $\chi^{(2)}_{xxx}(t,t+\tau)$ and $\chi^{(2)}_{xxx}(t,t+\tau,t+\tau)$ show signature of spinon decay as well. These results are qualitatively consistent with the physical picture presented in Ref.~\onlinecite{Wan2019}.

A main advantage of the methodology employed in this work is its versatility. It can be adapted to other one-dimensional spin models with little modification. In addition to the Ising ladder model, we have also calculated the two-dimensional coherent spectrum of a perturbed Ising chain, which has been analyzed recently by using the many-body perturbation theory \cite{Hart2022}. It's Hamiltonian is given by $H = H_0 + V$. $H_0$ describes an (extended) quantum Ising chain:
\begin{subequations}
\begin{align}
H_0 = \sum_n (-J\sigma^z_n\sigma^z_{n+1}-J_2\sigma^z_n\sigma^x_{n+1}\sigma^z_{n+2}-h\sigma^x_n),
\end{align}
where the $J_2$ term introduces a second-neighbor hopping of the spinons. $V$ introduces interactions between the spinons and thus spoils the integrability:
\begin{align}
V = \lambda\sum_n \sigma^x_n \sigma^x_{n+1} + \sigma^y_n \sigma^y_{n+1}.
\end{align}
\end{subequations}
Owing to the interaction, a spinon can decay into the continuum of three spinons. Fig.~\ref{fig:perturbed_model} shows the nonlinear magnetic susceptibility $\chi^{(3)}_{xxxx}(t,t,t+\tau)$ and the corresponding two-dimensional coherent spectrum of this model, where we set $J=1$, $h=0.4$, $J_2 = -0.48$, and $\lambda = 0.2$. From the time-domain data, we observe a clear spin echo signal as well as a persistent oscillation, periodic in both $t$ and $\tau$. We believe this oscillation arises from a possible anti-bound state of spinons. Accordingly, in the two-dimensional spectrum, we see the an additional feature in proximity to the spinon echo peak. This primitive calculation shows that interactions can give rise to interesting effects beyond the spinon decay, which worth further investigation.

Our work represents a first pass at the numerical simulation of the two-dimensional coherent spectra of interacting spinons. Based on the iTEBD algorithm, the accessible simulation is limited by the growth of entanglement entropy at late time. For the Ising ladder model, using a moderately large bond dimension $D = 1200$, the accessible simulation time is $t+\tau\approx 30/(J+h)$. As a result, we are unable to resolve the intrinsic line width of the spinon echo peak or to carry out any quantitative analysis. Therefore, a pressing challenge is to reach longer simulation times. On this front, we think it be useful to explore novel matrix-product based methods such as the folding method \cite{Banuls2009,Huang2014} and the density matrix truncation \cite{White2018} that might circumvent the entanglement entropy barrier.

In the phenomenological theory, the dynamics of the spinon pairs is mapped to that of an ensemble of two-level systems. The theory introduces two phenomenological time scales, the $T_1$ and $T_2$ times to describe respectively the de-population and the decoherence of an excited pair of spinons, and shows that these time scales can be read off from the two-dimensional coherent spectrum. We are yet unable to extract these time scales due to the limited simulation time. The recent many-body perturbation theory calculation made a first step in this direction \cite{Hart2022}. We think it would be interesting to use the numerical results, which are exact in early times, to benchmark the analytical calculations.

While in preparation of this manuscript, we become aware of Ref.~\onlinecite{Sim2022}, which reported the numerical calculation of the second order magnetic susceptibility of the quantum Ising chain subject to an longitudinal magnetic field by using the iTEBD algorithm. In addition to the different focuses, the methodologies are also different --- Ref.~\onlinecite{Sim2022} calculates the nonlinear susceptibilities directly using the Kubo formula, and performs a summation over lattice sites to obtain the optical nonlinear responses. In this work, we use the subtraction method to obtain the optical nonlinear responses, which avoids the summation. It might be interesting to compare the performances of these two methods in the future.

\begin{acknowledgements}
This work is supported by the National Natural Science Foundation of China (Grant No.~11974396) and by the Chinese Academy of Sciences through the Strategic Priority Research Program (Grant No.~XDB33020300), the Project for Young Scientists in Basic Research (Grant No. YSBR-059), and the Youth Innovation Promotion Association (Grant No. 2021004).
\end{acknowledgements}

\bibliography{interacting_spinon_nonlinear.bib}
\end{document}